\documentclass[ag]{copernicus}

\begin{document}

\title{Magnetic guide field generation in thin collisionless current sheets}

\author[1]{R. A. Treumann\thanks{Visiting the International Space Science Institute, Bern, Switzerland}}
\author[2]{R. Nakamura}
\author[2]{W. Baumjohann}

\affil[1]{Department of Geophysics and Environmental Sciences, Munich University, Munich, Germany}
\affil[2]{Space Research Institute, Austrian Academy of Sciences, Graz, Austria}

\runningtitle{Weibel fields in thin current layers}

\runningauthor{R.A. Treumann, R. Nakamura, and W.Baumjohann}

\correspondence{R.A.Treumann\\ (rudolf.treumann@geophysik.uni-muenchen.de)}

\received{ }
\revised{ }
\accepted{ }
\published{ }


\firstpage{1}

\maketitle

\begin{abstract} In thin ($\Delta<$ few $\lambda_i$) collisionless current sheets in a space plasma like the magnetospheric tail or magnetopause current layer, magnetic fields can grow {from  thermal fluctuation level by the action of the non-magnetic Weibel instability \citep{weibel1959}. }The instability is driven by the counter-streaming electron inflow from the `ion diffusion' (ion inertial Hall) region into the inner current (electron inertial) region from where the ambient magnetic fields are excluded when released by the inflowing electrons which become non-magnetic on scales smaller than the electron gyroradius and $<$ few $\lambda_e$. It is shown that under magnetospheric tail conditions it takes $\sim$ 20-40 e-folding times ($\sim$ 10-20 s) for the Weibel field to reach observable amplitudes $|{\bf b}_{\,\rm W}|\sim 1$ nT. In counter-streaming inflows these fields are predominantly of guide field type. This is of interest in magnetic guide field reconnection. Guide fields are known to possibly providing the conditions required for the onset of bursty reconnection \citep {drake2006,pritchett2005a,pritchett2006a,cassak2007}.  In non-symmetric inflows the Weibel field might itself evolve a component normal to the current sheet which could also contribute to reconnection onset.

\end{abstract}

\introduction
{In this communication we investigate the self-consistent generation of a so-called magnetic guide field in a thin collisionless current layer that initially lacks the presence of any guide field. For this we render responsible the Weibel instability \citep {weibel1959}.

Guide fields are believed -- and have been shown by numerical simulations \citep[see e.g.][and others] {drake2006,pritchett2005a,pritchett2006a,cassak2007} --  to be of prime importance in collisionless reconnection. Their presence in a thin collisionless current sheet is not evident. In a collisionless plasma the state of magnetisation is determined by the history of the plasma. External magnetic fields are unable to penetrate it; they have to be present in the plasma from the very beginning.  It is thus of vital interest in reconnection to understand whether guide fields can arise in collisionless current sheets. }

Space observations {\it in situ} \citep {fujimoto1997,nagai1998,oieroset2001,runov2003,nakamura2006,vaivads2004} and kinetic numerical simulations \citep {ramos2002,drake2003,drake2005} unambiguously prove that magnetic reconnection in a collisionless space plasma proceeds under the following two conditions:
\begin{itemize}
 \item{ that the current sheet that separates the oppositely directed (`anti-parallel') magnetic fields $\pm{\bf B}$ to both sides of the current becomes `thin enough'. Under `thin enough' it is understood  that the effective half-widths $\frac{1}{2}\Delta\lesssim\lambda_i$ of the sheets fall (approximately) below the ion inertial scale-length $\lambda_i=c/\omega_{i}$ (with $c$ velocity of light, $\omega_{i}=e(N/\epsilon_0m_i)^\frac{1}{2}$ ion plasma frequency, $e$ elementary charge, $N$ plasma number density, $m_i$ ion mass). The real limit on the width is not precisely known.   {Observations \citep{nakamura2002,vaivads2004,nakamura2006,baumjohann2007} suggest that it can reach values $\Delta\lesssim 4\lambda_i$. This condition is commonly  \citep[for example in][]{retino2007,sund2007} taken as sole indication that one is dealing with collisionless reconnection.}}
 \item{that plasma flows in into the sheet from both sides along the normal to the current at a finite (though low) velocity $\pm V_n$, explicitly considered only in treatments of  `driven reconnection'. Numerical simulations usually take care of this condition by starting the simulation with a prescribed reconnection configuration either assuming an initially present 
     X-point or by locally imposing a temporary parallel electric field or finite resistance for sufficiently long time in order to ignite reconnection.}
\end{itemize}
The first condition implies that ions in the sheet become inertia-dominated and thus are demagnetised, while electrons remain magnetised. Since magnetised electrons are tied to the magnetic field,  the magnetic field in the ion inertial region is carried along by the electrons causing Hall current flow \citep[as was realised  first by][]{sonnerup1979}.

From the second condition it is clear that reconnection cannot proceed continuously on time scales shorter than the inflow time $\tau_{\rm in} \simeq \Delta/V_n\sim {\rm few}\,\lambda_i/V_n$.   {If reconnection turns out to be faster, it will necessarily be non-stationary and probably pulsed.}

These just represent {\it necessary} conditions still being insufficient to describe the onset of reconnection.   {A collisionless mechanism is missing so far that either demonstrates, in which way the electrons become scattered away from the magnetic field in order for letting the oppositely directed magnetic field components slide from the electrons and reconnect, or that forces reconnection to occur in some other way. Prime attention in this respect is attributed to the action of guide fields, i.e., a magnetic field component in the direction of current flow.

Magnetic guide fields have two implications which in a thin current sheet with unmagnetised ions affect basically only the electrons:
\begin{itemize}
\item{that the centre of the current sheet is not free of magnetic fields, as would be the case in the Harris current sheet model, and}
\item{that the sheet current ${\bf J}$ possesses a field-aligned component $J_\|$ which, when exceeding a certain threshold, undergoes instability and may cause either anomalous dissipation via generation of anomalous resistance or viscosity, or produces localised wave fields like solitons and (electron) holes. Both resistance and guide-field-aligned electric fields violate the electron frozen-in condition and thus contribute to (localised) reconnection.}
\end{itemize}}

\begin{figure*}[t]
\centerline{{\includegraphics[width=0.8\textwidth,clip]{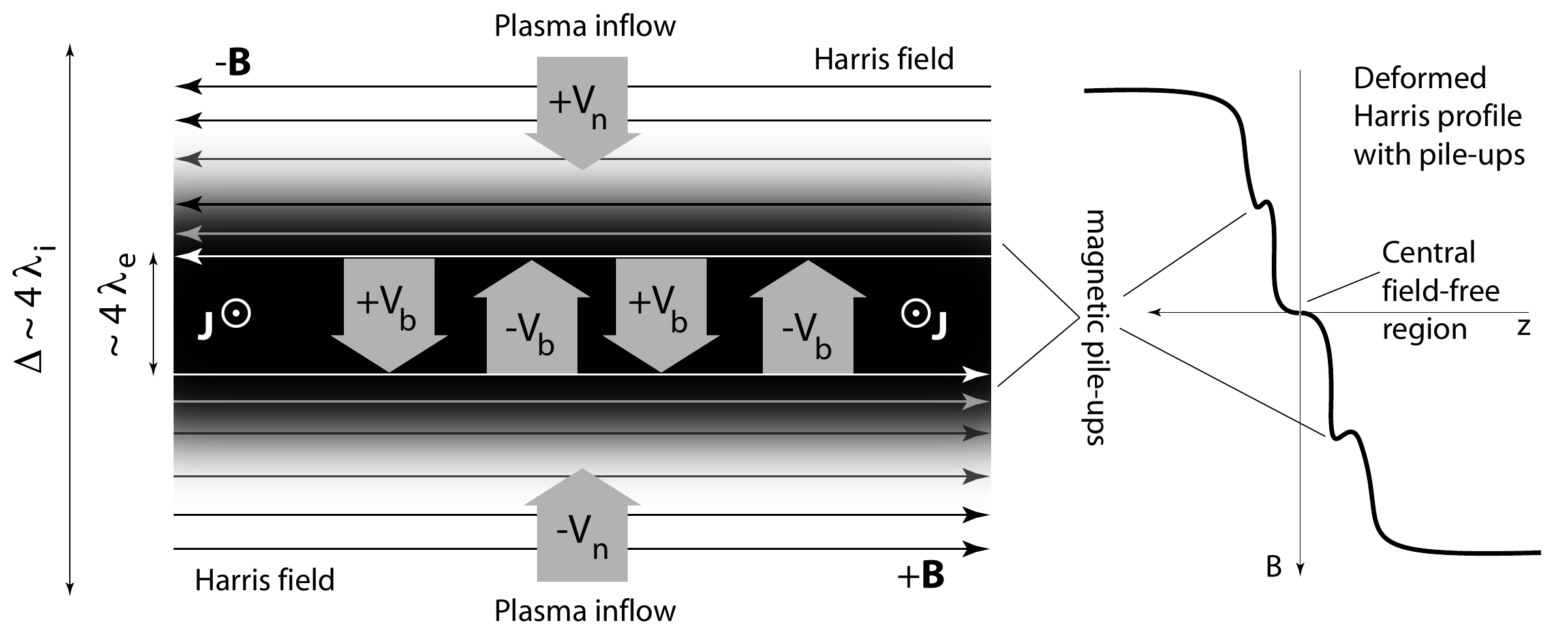}}}

\caption{Sketch of a homogeneous thin collisionless current sheet with plasma inflow from both sides and central electron inertial region. The magnetic field piles up at the boundary of the electron inertial region giving rise to a broad (of size of a few electron inertial lengths $\lambda_e=c/\omega_{\,e}$) field-free central current layer of sheet current ${\bf J}$ which is crossed by two (symmetric counter-streaming) electron flows which are Weibel unstable. On the right the pile-up deformed Harris profile is shown. It exhibits two shoulders in the regions where the magnetic field piles up at the boundaries of the broad field-free central electron inertial region, where the Weibel instability may generate magnetic guide fields. }\vspace{-0.3cm}\label{weibel1}
\end{figure*}

Quite generally, any guide-field-free current sheet model that scatters electrons away from the magnetic field turns out to be in trouble because of the following reason:

Assume that the electrons have transported the magnetic field from both sides some distance across the ion inertial region into the current sheet. During this transport the magnetic field lines have become bent locally thereby {\it enhancing} the magnetic tension. However, magnetic fields are massless and therefore have no inertia. Once the electrons enter the electron inertial region in the centre of the sheet, they demagnetise and release the magnetic field lines.  When this happens, the field lines react to the magnetic stress and will, hence, {\it rebound to their initial straight magnetic field configuration,} stopping moving further in for meeting their oppositely directed partner-field lines and merging with them. This behaviour will keep the internal section of the current sheet free of magnetic fields.   {(The stationary kinetic Harris equilibrium does not cover this effect; it just prescribes a $\tanh$-shape for the magnetic field and does not distinguish between ion and electron inertial regions.)}

Clearly, the rebounding field lines will again become loaded with electrons when snapping back into the `ion diffusion'   {Hall-current} region being unable to   {completely} regain their initial positions.   {Thus, magnetic field lines accumulate near the boundary of the electron inertial region until -- for continuous further plasma inflow (e.g. in driven reconnection scenarios) -- the accumulating external pressure becomes strong enough to push the newly created magnetic wall further in. }

In this respect it is of interest that very few of the magnetic profiles across a presumably reconnecting current sheet at the Earth's magnetopause \citep[cf., e.g.,][]{haaland2004,haaland2005,paschmann2005} or also in the magnetospheric plasma sheet \citep{runov2003} exhibit the canonical Harris $\tanh$-shapes; most of them show the evolution of either a magnetic plateau across the current sheet centre or regions along the profile where the magnetic field spatially undulates. The latter behaviour is conventionally attributed to waves passing along the boundary \citep{paschmann2005} referring to related changes in the boundary normal.   {This seems reasonable; part of it might, however, also be explained quite naturally by the above effect, in particular when there is tangential flow along the current layer -- as is the case at the magnetopause.}

Reconnection takes place   only, when the oppositely directed magnetic field lines are brought into direct physical contact to (partially) cancel each other, i.e. coming so close that the field-lines touch each other.\footnote{A precise formulation of this condition requires the definition of a field line cross section which cannot be given in classical plasma physics but requires a quantum electro-dynamical approach (R.A.Treumann, The quantum picture of reconnection, in preparation, 2010).} This, however, requires some mechanism that acts even deep inside the electron inertial region on scales   {$\lesssim \,{\rm few}\,\lambda_e=c/\omega_{e}\sim$ few km ($\lesssim$ 10 km in Earth's tail plasma sheet and even $\sim$ 1 km in the magnetopause region; with $\omega_{e}=e\sqrt{N/\epsilon_0m_e}\approx 56.41\sqrt{N}$ krad/s angular electron plasma frequency, $m_e$ electron mass, and $N$ is in cm$^{-3}$)} transporting the magnetic field further in. It is not known in which way non-magnetic electrons could cause further inward transport of the magnetic field.   {In quantum solid state physics such a transport is related to the celebrated integer and fractional quantum Halls effects \citep[see, e.g.,][for an almost complete account]{ezawa2000}. However, due to their high temperatures $T\gg T_{\rm F}$ (with $T_{\rm F}$ Fermi temperature) space plasmas are classical, and -- presumably -- no similar possibility opens up here. }

  {Thus the question arises whether a collisionless current sheet can by itself cause a guide field to evolve in its centre which circumvents the application of an external guide field, re-magnetises the demagnetised electrons in the centre of the current sheet and thus allows transport of the lobe magnetic field into the centre of the current sheet, contact of oppositely directed fields, and finally makes onset of reconnection possible.}

In this Letter we demonstrate that the Weibel instability \citep{weibel1959}   {is capable of doing this sufficiently fast. The newly generated Weibel guide-field could become comparably strong, a non-negligible fraction of the undisturbed external field. As expected it is directed along the current layer and thus indeed plays the role of a guide field, suggesting that under certain conditions collisionless narrow current sheets self-consistently generate weak guide fields} of strength $B_g/B_0<1$ where $B_0$ is the magnetic field strength in the inflow region. The current sheet then becomes   {locally capable of producing conditions under which reconnection takes place.  }

\section{Weibel scenario}
{The Weibel instability \citep{weibel1959} is known to produce stationary magnetic fields under conditions when the plasma exhibits certain anisotropies in flow and/or temperature. It is thus of interest to investigate the conditions for its excitation in a collisionless thin current sheet. This instability is driven either by electrons or ions with the ion instability being much weaker than the electron instability.

The original proposal by  \citet{weibel1959} referred to a temperature anisotropy in the unmagnetised electron distribution providing the free energy for a stationary (very low frequency $\omega\sim 0$) magnetic instability. }Various variants of this instability have been investigated in the past two decades \citep{yoon1987,yoon2007,medvedev1999,califano2002,silva2002,fonseca2003,achterberg2007} both in the non-relativistic and relativistic domains, unmagnetised and magnetised, cold and hot plasmas, and for temperature as well as beam instabilities, with the beams effectively faking a temperature anisotropy. Application of this instability was mostly intended in either laser (inertial) plasma fusion or the violent conditions present in astrophysical systems. As we will show below, reconnection provides one of the simplest and most interesting classical applications of the Weibel instability.

Figure \ref{weibel1} sketches the collisionless reconnection site, conventionally called the `ion-diffusion region' (even though there is no diffusion). The ions just become non-magnetic here. The frozen-in electrons  have sufficient momentum to continue their inward ${\bf E\times B}$-drift motion transporting the magnetic field to the centre of the thin current sheet. Close to the centre the electrons become demagnetised and release the magnetic field which snaps back as explained above. In the narrow electron inertial region of size of few $\lambda_e$ perpendicular to the current sheet the collisionless inert electrons still maintain their inward velocity ${\bf V}=\pm V_b{\hat {\bf z}}$ (with $V_b\lesssim V_n$) on both sides of the current sheet. Since the electrons are completely collisionless, the two flows pass across each other without (direct) interaction thereby realising a {\it non-magnetic} counter-streaming electron beam configuration which according to \citet{weibel1959} may become electromagnetically unstable.

The Weibel instability generates a {\it non-oscillating} ($\omega\sim0$) transverse magnetic field ${\bf b}$ with ${\bf k}$-vector about perpendicular to the electron beams ($k_\perp\gg k_\|$, subscripts refer to the direction of ${\bf V}$).  In space plasma this instability is manifestly non-relativistic, and as usual one might suspect that it is too   {weak} to cause any susceptible effect. We will, however, show below that its effect is not negligible. Investigating instability we work in the fluid approximation of cold ($T_b\ll T_e$) beams of density $N_b$,  also for simplicity assuming that the plasma is cold as well, even though the centre of the reconnecting current layer contains a denser $N>N_b$ thermal plasma of temperature $T=T_e+T_i$ (to which we will return when estimating the thermal fluctuation level of the instability). Under  {\it unmagnetised} plasma conditions in the presence of beams the Weibel instability grows. So far it has been investigated mostly in the relativistic limit. In reconnection in space plasmas inclusion of relativistic effects is unnecessary. The inflow Mach numbers are small, and beam velocities $V_b\ll c$ are small as well.

The two symmetric counterstreaming electron beams in the central electron-inertial region have roughly same density $N_b$,   a condition that applies primarily to the magnetospheric tail-current sheet; at the magnetopause the beams are non-symmetric. At very low frequencies the electromagnetic dispersion relation factorises \citep[see also the more elaborate papers by][and many others where non-symmetric, thermal and kinetic corrections are given and also external magnetic fields are applied, with the expected result that the latter act stabilising on the evolution of the instability]{yoon1987,achterberg2007}.

In slab geometry applicable to the geomagnetic tail current sheet, the factor describing plane {\it electromagnetic} fluctuations of frequency $\omega\approx 0$ becomes
\begin{equation}
{\textsf D}_{xx}{\textsf D}_{zz}-|{\textsf D}_{xz}|^2=0
\end{equation}
where ${\textsf D}_{ij}$ are the components of the dispersion tensor ${\textsf D}(\omega, {\bf k})=k^2c^2{\textbf{\textsf I}}-{\mathbf{\epsilon}}(\omega,{\bf k})$ (with plasma dielectric tensor function ${\mathbf\epsilon}(\omega,{\mathbf k})$). Under the assumed symmetric conditions ${\textsf D}_{xz}\equiv 0$, and the dispersion relation simplifies to
\begin{equation}
{\textsf D}_{zz}=n^2-1+\sum_s\, ^{s\!}\chi_{\,zz}=0, \qquad n^2=k^2c^2/\omega^2
\end{equation}
where $n$ is the refraction index, ${\bf k}$ wave number, $\omega$ wave frequency, and $^{s\!}{\bf \chi}_{ij}$ the susceptibility tensor of species $s$   {the only surviving component of} which, in a symmetric electron/electron-beam plasma configuration, is given by
  {\begin{equation}
\chi_{\,zz}=\frac{k^2V_b^2}{\omega^2}\frac{\omega^2_b}{\omega^2}+\frac{\omega^2_e}{\omega^2}\left(1+\frac{m_e}{m_i}\right)^{\!\!\!-1}
\end{equation}}
The subscripts $e$ and $b$ indicate background and beam parameters, respectively, $\omega_b$ is the beam plasma frequency for symmetric beam density   {$N_\pm=N_b/2$}, and in the last term the (negligibly small) neutralising background-ion contribution to the plasma frequency is taken into account for correctness in the electron-to-ion mass ratio term $m_e/m_i$. When the background plasma is at rest, the `wave' becomes non-oscillating with $\omega=\pm i\gamma_{\,\rm W}$   {(otherwise when the plasma moves at velocity ${\bf V}_0$ the wave frequency will be Doppler shifted by the amount ${\bf k\cdot V}_0$, a case that may be realised under magnetopause conditions, there with ${\bf V}_0$ being the magnetosheath flow velocity tangential to the magnetopause)}.
\begin{figure}[t]
\centerline{{\includegraphics[width=0.4\textwidth,clip]{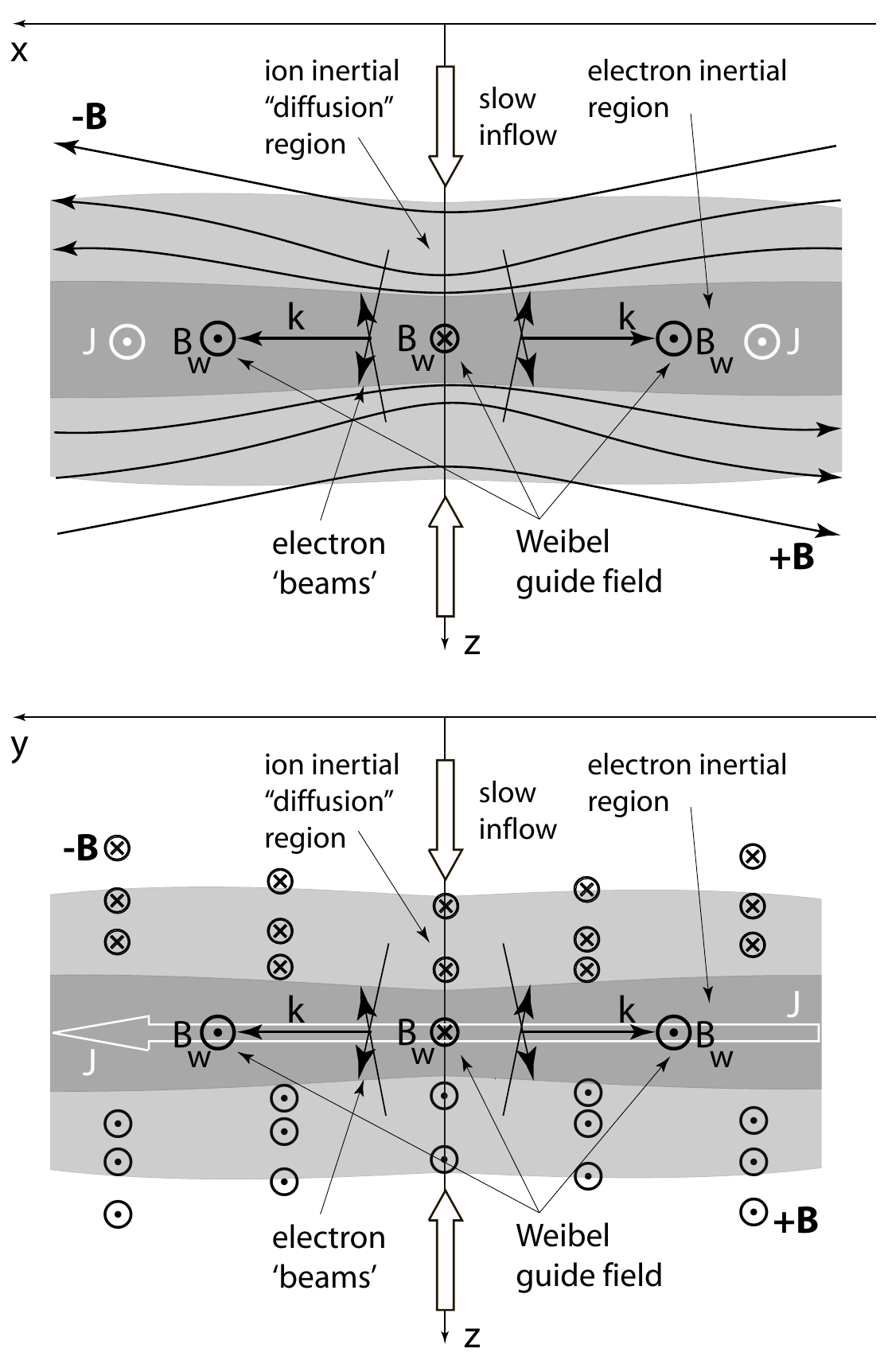}}}
\caption{Conditions inside a local electron inertial region. The external magnetic field accumulates at the boundary of the inertial region. Electron flow is perpendicular to the magnetic field. Inside the inertial region the electron components form two oblique beams. The Weibel instability generates a magnetic vortex of field ${\bf b}_{\,\rm W}$ with wavenumber ${\bf k}\perp {\bf V}_b$ perpendicular to the beam configuration. These vortices close in the current plane $(x,y)$ and have wavelengths of the order of $\sim \lambda_{eb}$. The sense of the magnetic vortices alternates along the current direction. Shown is the part of the lowest order vortex where the field is along the sheet current ${\bf J}$. At a phase of $\pi/2$ along the current the Weibel fields point along the ambient magnetic field causing wobbling of the current layer. Under non-symmetric beam conditions the wave vector and Weibel fields may become tilted against the current sheet. This could give rise to the local production of a normal magnetic field component $b_{z}$ which may possibly initiate reconnection.}\vspace{-0.3cm}\label{weibel2}
\end{figure}

Solving the above dispersion relation for $\gamma_{\,\rm W}>0$ yields the  growth rate  of the non-evanescent growing mode
\begin{equation}
\frac{\gamma_{\,\rm W}}{\omega_{\,b}}=\frac{V_b}{c}\left\{1+\frac{\omega_{b}^2}{k^2c^2}\left[1 +\frac{  {2}N}{N_b}\left(1+\frac{m_e}{m_i}\right)^{\!\!\!-1}\right]\right\}^{\!\!\!-\frac{1}{2}}
\end{equation}
Again, $N$ is the quasi-neutral background density. Maximum growth rates
\begin{equation}
\gamma_{\,\rm W, max}\approx \omega_{\,b}V_b/c
\end{equation}
are obtained when the second term in the braced expression becomes small, which is the case at relatively {\it short} wavelengths
\begin{equation}
k\lambda_{eb}\gg \left[1+(2N/N_b)\right]^\frac{1}{2} > 1
\end{equation}
The Weibel instability will thus lead to comparably   {small-scale} magnetic structures populating the beam-electron inertial range of size $\lambda\sim 2\pi\lambda_{eb}=2\pi \,c/\omega_{\,b}$.   {This condition of maximum growth seems to restrict the range of unstable wavelengths to very short scales. However, since the beam density is substantially less than the density of the ambient plasma in the centre of the current sheet, one has $\lambda_{eb}>\lambda_e$, the condition that $k\lesssim\lambda_e$ is thus not in contradiction with the condition of maximum growth.

The unstably generated magnetic structures are not purely transverse, however,} because $kc/\omega > 1$. They are of mixed polarisation, containing a longitudinal electric `wave' field component ${\bf e}_{\ell}  \parallel  {\bf k}$ which in the symmetric beam case is small, however. The main transverse electric field component ${\bf e}_{\,\rm T}$ is along the electron beams. Since the `wave' magnetic field ${\bf b}$ satisfies the solenoidal condition ${\bf b}\cdot{\bf k}=0$, ${\bf k}\perp{\bf V}_b$, and $\omega{\,\bf b}={\bf k\times e}_{\,\rm T}$, the Weibel magnetic field lies in the plane that is perpendicular to the plasma inflow into the current sheet (cf. Figure \ref{weibel2}).

In an extended two-dimensional current sheet the Weibel magnetic field is directed either parallel or anti-parallel to the sheet current   {${\bf J}= J{\bf\hat y}$} while being confined to the electron inertial zone around the non-magnetic centre of the current sheet.   {Such a field has the required properties of a guide field. Since it cannot be stronger than the magnetic field $B_0$ in the external inflow region, the Weibel instability generates weak guide fields} satisfying $b_{\,\rm W}/B_0<1$. The above smallness condition on its wavelength implies that the Weibel guide field forms a   {comparably short-scale wavy magnetic structure along the sheet current, thereby structuring the current sheet magnetically in the direction of the current flow.} In addition it has a transverse electric field ${\bf e}_{\rm T} || {\bf V}_b$ component along the electron beam inflow direction that is confined to the electron inertial region, while the longitudinal electric field is in the direction ${\bf e}_\ell\perp{\bf V}_b$.

The case that the inflow is not homogeneous either along the current or transverse to it is of particular interest.   {Then} the wave vector ${\bf k}_\perp$ rotates in the ($x,y$)-plane perpendicular to the inflowing electron beam. The condition that the magnetic field be free of divergence forces the Weibel magnetic field to form closed magnetic vortices in this plane. Such a magnetic vortex causes a magnetic asymmetry along the sheet current by periodically amplifying and weakening the external magnetic field on one side of (above or below) the current layer. This produces a spatial oscillation around the symmetry plane of the current layer of wavelength of the Weibel magnetic vortices.

\section{Thermal Weibel level}
So far we have been dealing with linear growth of the Weibel instability. In order to obtain its saturation level one needs to investigate the nonlinear evolution of the Weibel instability. Here we ask for how long it takes the Weibel instability to grow {from thermal fluctuation level until} reaching any measurable magnetic field strength.

To estimate the thermal Weibel level we refer to thermal fluctuation theory of non-magnetic plasmas \citep{sitenko1967,sitenko1982,akh1975, yoon2007a}. (Working in the Weibel low frequency limit implies that in their general kinetic expressions for the field fluctuations the variable $z\sim\omega$ must ultimately be set to zero.)   {Recently the Weibel thermal level in an isotropic thermal has been calculated explicitly \citep {yoon2007a}. There a very simple expression for the wave number dependence of the spectral energy density of the magnetic field fluctuations has been obtained as
\begin{equation}
\langle |{\bf b}(0,{\bf k})|^2\rangle_{\rm th} \propto 1/k^3\lambda_D^3, \quad \lambda_D = {\rm  Debye\ length}
\end{equation}}
According to this expression the magnetic fluctuations maximise at long wave lengths $k\to 0$. In order to avoid and understand this effect, we assume that the thermal background plasma in the current sheet has a weak thermal anisotropy $T_\perp\neq T_\|$. The direction of this anisotropy is arbitrary. Hence, for simplicity we assume that the $||$ direction refers to the direction of the beam (perpendicular to the current sheet), while $\perp$ refers to the direction of current flow. This choice is not unreasonable in view of the possibility that the mere presence of the current may correspond to a thermal anisotropy. In equilibrium the background plasma then obeys a bi-Maxwellian velocity distribution
\begin{equation}
f_0(v_\|,v_\perp)=\frac{(m_e/2\pi)^\frac{3}{2}}{T_\perp \sqrt{T_\|}}\!\exp\left[ -\left(\frac{m_ev_\perp^2}{2T_\perp}+\frac{m_ev_\|^2}{2T_\|}\right)\right]
\end{equation}
In the classical limit of  thermal fluctuation theory the magnetic spectral energy density $\langle b_ib_j\rangle_{k\omega}$ in an isothermal plasma is determined from Eq. (2.52) in \citet {sitenko1967} as
\begin{equation}
\langle b_ib_j\rangle_{k\omega}=\frac{\mu_0Tn^2}{\omega}\left(\delta_{ij}-\frac{k_ik_j}{k^2}\right)\frac{{\rm Im}\,\epsilon_\perp(\omega,{\bf k})}{|\epsilon_\perp-n^2|^2}
\end{equation}
In a thermally anisotropic plasma one must replace the temperature $T$ by the effective temperature $T_\perp T_\|/(T_\perp+T_\|)$ in this expression. Moreover, the transverse dielectric function $\epsilon_\perp(\omega,{\bf k})$ in the anisotropic case is given by
\begin{equation}
\epsilon_\perp(\omega,{\bf k})=1-\frac{\omega_e^2}{\omega^2}\left\{1-\frac{T_\perp}{T_\|}\left[1-\Phi(z)+i\pi^\frac{1}{2}z{\rm e}^{-z^2}\right]\right\}
\end{equation}
Here $z=\omega/\sqrt{2}kv_{e\|}$ is a variable that vanishes with $\omega\to 0$, $n^2\equiv (kc/\omega)^2=\epsilon_\perp$ is the refraction index of transverse fluctuations, and the real function $\Phi(z)\approx 2z^2$ for $z\ll 1$.

With the help of these expressions the spectral energy density $\langle |{\bf b}|^2\rangle_{k\omega}$ can be brought into the form
\begin{equation}
\langle |{\bf b}|^2\rangle_{k\omega} =\frac{2\mu_0}{\omega_e}\frac{\sqrt{\pi/2}(c/v_{e\|})T_\|[\theta/(1+1/\theta)] \kappa {\rm e}^{-z^2}}{[\kappa^2-{\tilde\omega}^2+1-\theta(1-\Phi)]^2+\pi\theta^2z^2{\rm e}^{-2z^2}}
\end{equation}
where we defined $\kappa\equiv k\lambda_e, {\tilde\omega}\equiv\omega/\omega_e,\theta\equiv T_\perp/T_\|$, and $\lambda_e=c/\omega_e$ is the background plasma electron inertial scale. This equation holds for small frequencies $\omega\simeq 0$. Moreover, one now realises that for fixed $\omega=$ const and small $k$, using $\Phi(z)\sim 1+1/2z^2 + \cdots$ asymptotically for $z\gg1$, the spectral energy density becomes
\begin{equation}
\langle |{\bf b}|^2\rangle_{k\omega} \propto \frac{\kappa {\rm e}^{-z^2}}{[\kappa^2-{\tilde\omega}^2+1+\theta /2z^2]^2+\pi\theta^2z^2{\rm e}^{-2z^2}}
\end{equation}
which, as required, vanishes exponentially in the limit $\kappa\sim k\to 0$, which holds for all frequencies $\omega$.

The spectral energy density of the Weibel field is obtained in the limit $z=\omega=0$, which yields
\begin{equation}
\langle |{\bf b}|^2\rangle_{k0} =\frac{2\mu_0}{\omega_e}\frac{\sqrt{\pi/2}(c/v_{e\|})T_\|[\theta/(1+1/\theta)] \kappa}{[\kappa^2+1-\theta]^2}
\end{equation}
The isotropic case $\theta=1$ reproduces the $k^{-3}$ scaling with wave number which diverges at $k\to 0$.
Retaining the temperature anisotropy the spectral energy density of Weibel fluctuations vanishes both for $\kappa\to 0$ and $\kappa\to\infty$. It, hence, has a maximum for $\theta<1$ (i.e. $T_\perp<T_\|$) at
\begin{equation}
\kappa_m\equiv k_m\lambda_e=\sqrt{\frac{1}{3}\left(1-\theta\right)}
\end{equation}
The value of this maximum is
\begin{equation}
\langle |{\bf b}|^2\rangle_{k_m0} =\frac{3\mu_0}{8\omega_e}\frac{\sqrt{\pi/2}(c/v_{e\|})T_\|[\theta/(1+1/\theta)] }{(1-\theta)}
\end{equation}
which, for small anisotropies simplifies to become
\begin{equation}
\langle |{\bf b}|^2\rangle_{k_m0} \approx \frac{3\mu_0 T_\perp}{8\omega_e(1-\theta^2)} \left(\frac{\pi}{2}\right)^{\!\!\!\frac{1}{2}}\! \!\frac{c}{v_{e \| } }, \qquad \theta < 1
\end{equation}
The condition $T_\perp<T_\|$ corresponds to Weibel equilibrium. In the opposite case $T_\perp>T_\|$ the thermal fluctuations explode at wavenumber
\begin{equation}
\kappa_\theta=+\sqrt{\theta-1}
\end{equation}
which indicates that in the vicinity of $\kappa_\theta$ the plasma is unstable with respect to the ordinary anisotropic Weibel instability. The unstable fluctuation wavenumber $\kappa_\theta$ becomes small and the fluctuations long-wavelength for small anisotropies $\theta\gtrsim 1$. Then the plasma spontaneously pumps energy into the zero-frequency magnetic fluctuations while leaving the thermal equilibrium state. This would be the case when the sheet current generates sufficient thermal anisotropy to spontaneously produce magnetic field fluctuations in the direction perpendicular to the current; once this happens, a selfconsistent magnetic field component will be created in the $z$-direction which could cause set-on of reconnection. Here, this particular interesting case will not be further considered.

In a current sheet of density $N\sim10^6$ m$^{-3}$, temperature $T\sim$ 0.1 keV which implies  electron inertial  and Debye lengths $\lambda_e\approx 6$ km and $\lambda_D\approx 7.5$ m respectively, presumably corresponding to conditions in the near-Earth magnetotail current sheet, the maximum Weibel spectral energy density becomes
\begin{equation}
\langle |{\bf b}|^2\rangle_{k_m0} \approx 3\times 10^{-26} \frac{T_{\rm eV}}{\sqrt{N_{\rm cm^{-3}}}} \,\, \frac{\rm V^{\,2}\,s^3}{\rm m}
\end{equation}
where $T_{\rm eV}$ is the background plasma electron temperature in eV, and $N_{\rm cm^{-3}}$ is the plasma density in cm$^{-3}$.

\section{Growth time in the magnetotail current layer}
We are interested in the time required for the Weibel instability to grow to measurable guide magnetic field values under conditions in the magnetotail reconnection region. Assuming an electron `beam' density $N_b\approx 10^{\,5}$ m$^{-3}$ corresponding to $N_b/N\sim 0.1$,  beam velocity $V_b\sim 30$ km\,s$^{-1}$, angular beam size $\Delta\alpha\lesssim 10^\circ$, the maximum growth rate of the Weibel instability in the tail current sheet becomes
\begin{equation}
\gamma_{\, {\rm W,max}}\lesssim 2 \,\,\,\, {\rm s}^{-1}
\end{equation}

The Weibel instability picks up the thermal spectral energy density at its wave number of maximum growth $k_{\rm W,max}^2\gtrsim [1+2N/N_b]/\lambda_{eb}^2$ which it amplifies. With the above numbers the Weibel wavenumber in the magnetotail current sheet becomes
\begin{equation}
k_{\rm W,max}\lambda_{eb}\gtrsim \sqrt{5}=2.24
\end{equation}
which corresponds to a maximum wavelength of $\lambda_{\rm W, max}< 2.8\, \lambda_{eb}$ in the magnetotail.

The thermal fluctuation energy density at maximum growing wave number is
\begin{equation}
\langle |{\bf b}|^2\rangle_{k_{\rm W}0} \lesssim\frac{2\mu_0}{\omega_e}\frac{\sqrt{\pi/2}(c/v_{e\|})T_\|[\theta/(1+1/\theta)] \kappa_{\rm W}}{[\kappa^2_{\rm W}+1-\theta]^2}
\end{equation}
where $\kappa_{\rm W}=k_{\rm W,max}\lambda_e\simeq(\lambda_e/\lambda_{eb}) [1+2N/N_b]^\frac{1}{2}>0$. In the isotropic case this simplifies to
\begin{equation}
\langle |{\bf b}|^2\rangle_{k_{\rm W}0}\lesssim\frac{\mu_0}{\omega_e}\sqrt\frac{\pi}{2}\frac{c}{v_{e}}\frac{T }{\kappa^3_{\rm W}}\approx\, 1.4\times 10^{-29} \ {\rm \frac{V{\,^2}\,s^3}{m}}
\end{equation}
Because the maximum Weibel wavenumber $\kappa_{\rm W}\neq0$ is finite, this expression is free of divergence.

We can now make use of the quasilinear instability growth prescription for the time evolution of the spectral energy density
\begin{equation}
\langle |{\bf b}(t, {\bf k},0)|^2\rangle\approx \left\langle|{\bf b}^2({\bf k},0)|\right\rangle_{\rm W,th}\exp (2\gamma_{\rm W} t)
\end{equation}
where on the left is the linearly growing spectral density at time $t$, and on the right the thermal level from where the instability starts growing. No nonlinear saturation effects are included at this time. An observable magnetic field in the magnetotail should be roughly of order $b\sim$\, 1 nT, a value which one may use to find the required growth time $\tau_{\rm W}$ from
\begin{equation}
\tau_{\rm W}\approx \frac{1}{2\gamma_{\rm W}}\ln\frac{\langle|{\bf b}^2({\bf k},\tau_{\rm W})|\rangle}{\langle|{\bf b}^2({\bf k},0)|\rangle_{\rm W,th}}
\end{equation}
Inserting for the thermal energy density $\left\langle|{\bf b}^2({\bf k},0)|\right\rangle_{\rm W,th}$ from $\langle |{\bf b}|^2\rangle_{k_{\rm W}0}$ and using the spectral energy density of a $b\sim$\,1\,nT magnetic field, $\langle b_{\rm 1 nT}^2 \rangle_{k0} \approx 4.3\times 10^{-12}\,\, {\rm V^2\,s^3 /m}$, the typical growth time to reach this magnetic field level becomes
\begin{equation}
\tau_{\rm W, max}\sim (10-20)\ {\rm s}
\end{equation}
corresponding to 20-40 e-folding times. This time is in the range of typical onset times of substorms and thus not in contradiction to the onset of reconnection in the magnetotail.

Since the maximum growing Weibel wave number is not precisely known (only a lower limit has been determined above), we can also use the maximum thermal fluctuation level at zero frequency which we have calculated for the anisotropic case.
With its help we find for the thermal magnetic pressure
\begin{eqnarray}
\frac{\left\langle |{\bf b}|^2\right\rangle_{\rm W,th}}{2\mu_0}&\simeq&\frac{2\pi\omega_e}{\mu_0\lambda_e^3} \left|\int {\rm d}{\tilde\omega}\,\kappa^2\,{\rm d}{{\kappa}}\,\delta({\tilde\omega})\delta(\kappa-\kappa_m)\langle |{\bf b}|^2\rangle_{\kappa 0}\right| \nonumber\\
&\approx&  1.7\times 10^{-27}(1-\theta) \quad {\rm J\,m}^{-3}
\end{eqnarray}
where we have used that, as has been argued above, the Weibel instability wave number $k\lambda_e\gtrsim 1$ is by order of magnitude comparable to the electron inertial scale $\lambda_e$. This thermal pressure corresponds to a thermal Weibel magnetic field strength
\begin{equation}
|\langle {\bf b}^2\rangle_{\rm W,th}|^\frac{1}{2}\sim 6.5\times10^{-17}\sqrt{1-\theta}\quad {\rm T}
\end{equation}
It is approximately this magnetic fluctuation level from where the Weibel instability in the current sheet starts to grow.   {Again, the quasilinear evolution of the spectral density can be used.
However, maximum growth is practically independent of  wave number $k$, and the same expression holds as well for the square of the maximum unstable Weibel-magnetic field.} If we assume that the instability can freely grow until reaching observable values before nonlinear effects come into play, the typical growth time for the magnetic field becoming observable is
\begin{equation}
\tau_{\,\rm W}\approx \frac{1}{2\gamma_{\rm W}}\ln \frac{\langle |{\bf b}(\tau_{\,\rm W})|^2\rangle}{\langle|{\bf b}|^2\rangle_{\rm W,th}}
\end{equation}
{We again demand that $|{\bf b}(\tau_{\,{\rm W}})|\approx 1\,$ nT in order to be observable in the magnetospheric tail plasma sheet. Since the calculation is anisotropic we assume the presence of a week anisotropy $1-\theta\sim 0.01$ only to again find a typical growth time of
\begin{equation}
\tau_{\,\rm W}(1\,\,{\rm nT})\approx {10}~{\rm s}
\end{equation}
which corresponds to $\sim\,$20 e-folding times. This is about the same time as that estimated above. Hence, the two approaches are mutually consistent. 

During this growth time the beam traverses a distance of roughly $\sim$ (600-1000) km $\gg \lambda_e$. {This implies that during the growth time of the Weibel field the electron beams cross the entire electron inertial region, i.e. the inner field-free region of current flow, thereby \textit{a posteriori} justifying our assumption of symmetric counterstreaming electron beams in the magnetotail.}} 

 \conclusions[Discussion and conclusions] The result of this investigation is that inside the magnetic field-free electron inertial region (of transverse size of a few $\lambda_e$) in the centres of thin -- possibly reconnecting -- current sheets, the inflow of electrons into the sheet from its two sides   {(in the geomagnetic tail from the lobes)} may well be capable of self-consistently generating a {\it weak magnetic guide} field via the non-magnetic Weibel instability.   {Given sufficient time, such weak guide fields will evolve in the very centre of the current sheet in the {\it electron} inertial region only}. (Similar weak quadrupolar guide fields are caused in the   {much more extended {\it ion} inertial zone by the Hall effect, but these fields are found {\it outside} the current sheet centre, do not penetrate into the electron inertial zone and, thus, are only indirectly involved into the reconnection process.})

  {The condition for the Weibel mechanism to work is that the ambient current sheet electrons have become nonmagnetic earlier than inert. This is the case that is realised in the centre of the magnetotail current sheet where the electron gyrodradius exceeds the electron inertial length. In terms of the electron thermal speed  this means that  initially
\begin{equation}
v_{e}>\sqrt{m_i/m_e}V_{A0}
\end{equation}
 where $V_{A0}$ is the Alfv\'en speed  outside the current layer based on the external field. For an initial Harris current layer one writes this as a condition on the vertical coordinate $z$
\begin{equation}
\frac{z}{\Delta}< \sinh^{-1}\left[\frac{v_{e}}{V_{A0}}\left(\frac{m_e}{m_i}\right)^{\!\frac{1}{2}}\right]
\end{equation}
In the magnetospheric tail we have $V_{A0}\sim 10^6$ m\,s$^{-1}$, and in the plasma sheet $T_e\sim 10^2$ eV which yields that $z<0.1\Delta$. In the inner region of the tail current this condition is clearly satisfied before entering the electron inertial region because $\lambda_e\sim 0.02\lambda_i$, and $\Delta\sim$ few $\lambda_i$.}

The Weibel-guide field $B_g$ is limited to be weaker than the ambient external magnetic field $B_0$. In the symmetric magnetospheric tail current sheet we find that it may reach up to $\lesssim10$\% of the ambient field in strength in a reasonably short linear e-folding time unless nonlinear effects set on earlier to saturate the field on a much lower level.   {Such nonlinear processes have not been investigated here. They can in principle be treated by assuming a stationary final state and calculating the thermal saturation level of the beam driven Weibel instability. This, however, makes sense only if it is assumed that the stationary state will be reached in a time shorter than the onset of reconnection. The two processes cannot be considered separately, since reconnection heavily depends on the presence of the (here self-consistently generated) guide field and will necessarily be fast, as has been shown by various numerical simulations that include the presence of even weak guide fields \citep[see e.g.,][]{pritchett2005a,cassak2007}.

Guide fields are important in the dynamics of the current sheet and in particular for reconnection. They have been observed in the geomagnetic tail in various cases \citep[see e.g.,][]{nakamura2008}.They  partially re-magnetise the electron plasma in the central current region. Pointing along the electric field that drives the current, they cause particle acceleration, which amplifies the current, generates energetic particles, and generates a number of secondary effects that affect the stability of the plasma in the current sheet. The role of guide fields in collisionless magnetic reconnection and the various effects it may cause have in the past decade been thoroughly investigated in numerical simulations both for guide fields perpendicular \citep[e.g.,][] {ricci2003,ricci2004,ricci2004a} and parallel \citep[e.g.,][] {pritchett2005a,pritchett2006a,cassak2007,egedal2008,le2009} to the sheet current. We may therefore refer to the published literature where the existence of guide field{s} has been imposed.

The interesting point we make here is that the Weibel instability provides a selfconsistent mechanism for producing such guide fields in the field free current region. Since the plasma there is collisionless on quite a long time scale, the existence of guide fields is a nontrivial problem. External fields can penetrate the plasma only on very long diffusion time scales, much longer than permitted by the observation of reconnection onset. Hence, a mechanism that generates internal guide fields is highly welcome. The non-relativistic Weibel instability may provide such a mechanism. The Weibel fields are comparably small-scale forming magnetic vortices of alternating polarity. On the scale of their wavelength, they cause a modulation of the magnetic field and consequently a spatial modulation of the current sheet. In the general case when the current sheet is not homogeneous, the guide fields may cause a three-dimensional structuring of the current sheet.

The calculations presented here applied to the case when the conditions are homogeneous along the current, such that ${\bf b\cdot {\hat z}}=0$. In the general case, when homogeneity is not given, the Weibel magnetic field may possibly develop a non-zero field component ${\bf b}_z=b_{n\rm W}{\bf\hat z}$ in the direction normal to the current sheet. Such a local normal magnetic field component is equivalent to a seed-X-point field and may ignite reconnection in a similar way as assumed for initial condition in numerical simulations.

  Proof of this conjecture requires a full three-dimensional investigation of the non-magnetic  Weibel instability under conditions when the current sheet is locally compressed (as sketched  in Figure \ref{weibel2}) in both $x$ and $y$ directions. If this conjecture turned out to be true, the local collisionless generation of Weibel magnetic fields in thin current sheets inside the electron inertial region $|z|\lesssim$ few $\lambda_e$ would provide a natural mechanism for initiating reconnection. The growth time of reconnection in such a case is the sum of the growth time $\tau_{\,\rm W}$ of the Weibel instability starting from thermal level $\langle {\bf b}^2\rangle_{\rm th}$ until the normal Weibel field component  $b_{n{\rm W}}$ becomes strong enough to modulate the ambient magnetic field, plus the well-explored growth time $\tau_{\,rec}$ of reconnection starting from an initial magnetic seed configuration. The latter is accessible through (and from existing) numerical simulations.

Finally, we remark that under anisotropic conditions with $T_\perp>T_\|$ the Weibel thermal level diverges, which indicates instability for a certain wavenumber $\kappa_\theta$ that depends on anisotropy $\theta>1$. Such conditions can, in principle, cause spontaneous growth of a magnetic field component $b_z$ perpendicular to the current sheet which would directly act as a seed field for spontaneous reconnection.




\begin{acknowledgements}
This research is part of a Visiting Scientist Programme at ISSI, Bern. Hospitality of the ISSI staff and directors is thankfully recognised.
\end{acknowledgements}










\end{document}